\documentclass{amsart}

\newtheorem{theorem}{Theorem}

\theoremstyle{definition}

\theoremstyle{remark}

\numberwithin{equation}{section}

\usepackage{graphicx}
\usepackage{amssymb}
\usepackage{amsmath}
\usepackage{mathtools}
\usepackage{verbatimbox}
\usepackage{algorithm}
\usepackage[noend]{algpseudocode}
\usepackage{scalerel}
\usepackage[symbol]{footmisc}
\usepackage[T1]{fontenc}
\usepackage{etoolbox}
\usepackage[symbol]{footmisc}

\usepackage{mathtools}

\usepackage{mathtools}
\DeclarePairedDelimiter\floor{\lfloor}{\rfloor}
\usepackage[T1]{fontenc}

\usepackage{hyperref}

\begin{document}

\title{Group Testing in the High Dilution Regime}

\author{Gabriel Arpino\textsuperscript{\textdagger}}
\address{ETH Z\"urich, Switzerland}
\curraddr{}
\email{garpino@ethz.ch}
\thanks{\textsuperscript{\textdagger} First two authors contributed equally, listed in alphabetical order}

\author{Nicolò Grometto\textsuperscript{\textdagger}}
\address{ETH Z\"urich, Switzerland}
\curraddr{}
\email{ngrometto@ethz.ch}
\thanks{}

\author{Afonso S. Bandeira}
\address{ETH Z\"urich, Switzerland}
\curraddr{}
\email{bandeira@math.ethz.ch}
\thanks{}



\date{\today}

\dedicatory{}

\begin{abstract}
Non-adaptive group testing refers to the problem of inferring a sparse set of defectives from a larger population using the minimum number of simultaneous pooled tests. Recent positive results for noiseless group testing have motivated the study of practical noise models, a prominent one being dilution noise. Under the dilution noise model, items in a test pool have a fixed probability of being independently diluted, meaning their contribution to a test does not take effect. In this setting, we investigate the number of tests required to achieve vanishing error probability with respect to existing algorithms and provide an algorithm-independent converse bound. In contrast to other noise models, we also encounter the interesting phenomenon that dilution noise on the resulting test outcomes can be offset by choosing a suitable noise-level-dependent Bernoulli test design, resulting in matching achievability and converse bounds up to order in the high noise regime.
\end{abstract}

\maketitle

\section{Introduction}
The goal of non-adaptive group testing is to identify the defective set $\mathcal{D}$ of cardinality $d$ among a much larger population of $n$ items using the least number of tests N. The aim is to exploit the sparsity in the problem by conducting simultaneous pooled tests, where outcomes are positive if and only if at least one defective item is included in the test. The problem has gained significant attention in many application areas such as communication protocols \cite{contention}, DNA sequencing \cite{PPR:PPR34919}, and most recently, COVID-19 testing \cite{mallapaty2020mathematical}. In recent years, noiseless group testing has become increasingly well understood in terms of its information-theoretic limits and the performance of decoding algorithms \cite{atia2012boolean, aldridgecap, johnson2018performance}.
Nevertheless, a significant gap exists in our understanding of noisy versions of the problem. Scarlett J. and Johnson O.~\cite{scarlett2020noisy} and Atia G. et al.~\cite{atia2012boolean} propose practical algorithms that are information theoretically optimal for certain regimes where the output of tests have i.i.d. error. 
This paves the way for understanding other practically relevant noise models. One such example arises in the context of medical and biological applications, where the contribution of a given defective item naturally lessens as the corresponding test pool increases in size, subject to the precision of testing instruments. The {dilution} noise model was firstly introduced in the work of Hwang F.K. \cite{hwang1976group} to model such cases, by assuming that defective items in a test are independently {diluted}, that is, their expected contribution to a test result may not take effect. The structure of this model is in contrast to the \textit{Z-channel} noise model where a test containing defective items, hence rightfully positive in the noiseless setting, has a fixed probability of being negative, regardless on how many defective items it contains. 

\noindent 
The information-theoretic properties of dilution noise versus noise on the test outcomes are yet to be understood. The subtle but significant change in noise description, together with the practical relevance of this noise model in applications, motivate the need for a rigorous analysis of both achievability and converse bounds for recovery of the defective set, noting that extra care is required in deriving such results in comparison to other noise models.
In this work, we analyse the dilution group testing problem in the case of i.i.d. Bernoulli test designs, where an item is independently included in a test with fixed probability $p$. An important feature of the dilution model is that the optimal choice of average test sparsity depends on the level of dilution noise. In particular, as we will see below, the required asymptotic number of tests for recovery is qualitatively different when choosing $p$ optimal for the noiseless setting versus $p$ depending on the dilution probability parameter. To be more precise, the fixed probability of an item being included in a test is a strictly increasing function of the noise level. In this paper, we characterize the asymptotic required number of tests in this setting.

\section{Related Work}
For noiseless non-adaptive group testing, state-of-the-art algorithms are known to achieve vanishing error probability with $N = \mathcal{O}(d \log{n})$ tests which is necessary and sufficient if vanishing probability is allowed \cite{scarlett2020noisy, sepmat}, motivated by classical results in error-free decoding outlining the best known bounds on the rate for testing pool designs \cite{r1, r2, r3}. Among these, the first practical group testing algorithm was $\mathtt{COMP}$, and its noisy variant $\mathtt{NCOMP}$. We choose this algorithm as the simplest example to illustrate the gains in achievability offered through noise-level-dependent test designs. \\ The dilution noise setting was first analysed by Atia et al. in \cite{atia2012boolean}, who derive order-optimal achievability bounds in the dilution noise setting (in noise-independent designs). They show that a maximum likelihood (ML) decoder recovers the defective set with vanishing error probability in the dilution noise model provided $\text{N} = \mathcal{O}(\frac{d \log{n}}{(1 - q)^2})$ where $q$ is the dilution noise probability. 
The work of Chan et al. in \cite{chan2011non} derives analogous achievability bounds for a Linear Programming (LP) decoder in the dilution noise setting, showing that this algorithm achieves order-optimal sample complexity under the action of dilution noise.\\
We highlight the squared dependence on $1-q$, for a dilution noise level $q$ in the current literature.
Although this work focuses on Bernoulli test designs, recent advances have introduced more sophisticated test designs such as the constant column weight design \cite{johnson2018performance} which would require a more intricate analysis, in the dilution noise setting. We  thus leave this further generalisation to future work.

\section{Problem Setup}\label{setup}
Consider a given population consisting of $n$ items labelled as $\{1,...,n\}$ and let $\mathcal{D} \subseteq  \{1,...,n\}$, of size $d$, denote the set of defective items in the population. We consider the \textit{sparse} regime, where $d = o(n)$. Let us assume a combinatorial prior on $\mathcal{D}$ \cite{aldridge2019group}, that is $\mathcal{D}$ is chosen uniformly at random over all $\binom{n}{d}$ possible subsets of $\{1,...,n\}$ of size $d$. The aim is to identify the defective items, using the minimum number of tests N. Each test consists of a sub-sample of items whose outcome (possibly noisy) is positive (indicated with 1) if and only if the current test pool contains a defective item. The test pools are formed prior to testing and each item is included with probability $\frac{\alpha}{d}$, independently, for a design parameter $\alpha$, noting that this guarantees a constant probability of having at least one defective in a given test pool. We encode the testing mechanism by considering indicator random variables $m_{i,j}$, for item $j$ being included in test $i$. We let $\text{M} = \{m_{(i,j)}\}_{i,j} \in \{0,1\}^{\text{N} \times n}$ be a Bernoulli design matrix, with each row vector specifying the item composition of the corresponding test pool and denote with $\mathbf{y} \in \{0,1\}^{\text{N}}$ the relative test results vector, whose i-th entry is obtained via an \texttt{OR} operation on the entries of the i-th row of M indexed by $\mathcal{D}$, indicated by 
\begin{align*}
\mathbf{y}_i = \underset{j \in \mathcal{D}}{\bigvee} m_{i,j}.
\end{align*}
A \textit{decoder} $\mathtt{A}$ is a function mapping pairs $(\text{M},\mathbf{y})$ to a subset of $\{1,...,n\}$, producing an estimate $\hat{\mathcal{D}}_\mathtt{A}$ for $\mathcal{D}$. The focus is on exact recovery, that is $\hat{\mathcal{D}}_{\mathtt{A}}$ computed by a chosen decoder $\mathtt{A}$ is successful if and only if $\hat{\mathcal{D}}_{\mathtt{A}} = \mathcal{D}$. We define the average error probability for a decoder $\mathtt{A}$ as $\text{P}^{(\text{e})}_{\mathtt{A}} = \mathbb{P}[\hat{\mathcal{D}_{\mathtt{A}}} \neq \mathcal{D}]$
where the randomness is over $\mathcal{D}$, the Bernoulli design matrix and the test results (in the noisy setting). 
We define the critical number of tests for decoder $\mathtt{A}$ as the minimum number of tests $\text{N}_{\mathtt{A}} = \text{N}_{\mathtt{A}}(n)$ for which, given any $\epsilon > 0$ and for $n$ sufficiently large, the decoder error probability $\text{P}_{\mathtt{A}}^{(\text{e})}$ is at most $\epsilon$. In what follows, we will derive an upper bound on $\text{N}_{\mathtt{NCOMP}}$ for $\mathtt{NCOMP}$ and a lower bound on $\text{N}_{\mathtt{A}}$ for any decoder $\mathtt{A}$ operating with a noise-level-dependent test design that takes the parametrisation $\alpha = \frac{\log{2}}{1-q}$, where $\log$ is used throughout to denote the natural logarithm. This parametrisation is chosen so as to keep the entropy of the resulting test vector constant (maximal) as $q$ varies. The chosen performance criterion for decoders is given by the number of bits of information on the defective set which are retrieved per test. This is measured via the rate \cite{baldassini2013adaptive}, defined as $\text{R}_{\mathtt{A}} = \frac{\log_2\binom{n}{d}}{\text{N}} \sim  \frac{d \log(n/d)}{\text{N}}$, where $\sim$ indicates scaling up to $1+o(1)$ factors, as $n\rightarrow \infty$. 
Under dilution noise, each defective item initially included in a given test gets diluted independently, hence not contributing to the test outcome. This is modelled by flipping each 1-entry in the design matrix to 0, independently, with probability $q$. The i-th test result is then given by
\begin{align*}
    \mathbf{y}_i = \underset{j \in \mathcal{D}}{\bigvee} \mathcal{Z} \left ( m_{(i,j)} \right )
\end{align*}
where $\mathcal{Z(\cdot)}$ indicates a Z-channel with flip probability $q$. Whilst the dilution model presents some structural similarities with Z-channel noise \cite{scarlett2020noisy}, in the former occurrence of noisy instances on the test results depends on the initial number of defectives included in a given test pool, but not in the latter, where noise merely acts on the test results.

\section{Achievability Bounds}
The noisy version of the $\mathtt{COMP}$ algorithm (\cite{aldridge2019group}, Section 3.4), $\mathtt{NCOMP}$, performs sequential decoding, by considering each item separately and declaring it defective if the number of positive tests containing said item is larger than a specified fraction of the total number of tests which contain it, specified via a design parameter $\Delta$. This algorithm is efficient as its run-time scales as a polynomial of the input size (n). The following theorem provides an achievability bound for $\mathtt{NCOMP}$ under dilution noise, above which $\text{P}^{(\text{e})}_{\mathtt{NCOMP}} \rightarrow 0$, as $n \rightarrow \infty$, in the case of a noise-level-dependent parameter $\alpha$. 

\begin{algorithm}
\caption{$\mathtt{NCOMP}$ for Dilution Noise}
\begin{algorithmic}[1]

\Procedure{\texttt{NCOMP}}{$\Delta, n, q$} 
\State \textbf{for} $i \in \{1,...,n\}$:
    \State $\mathcal{G}_i$: \#\{tests containing i\}
    \State $\mathcal{P}_i^+$: \#\{positive tests containing i\}
    \State Initialise: $\hat{\mathcal{D}}_{\mathtt{NCOMP}} \leftarrow \emptyset$
    \State \textbf{for} $i \in \left\{1,...,n\right\}$: 
        \If{$\mathcal{P}_i^+   \geq  \mathcal{G}_i  \cdot (1-q\cdot (1+\Delta))$}
        \State Include item i in $\hat{\mathcal{D}}_{\mathtt{NCOMP}}$
        \EndIf

    \State Return $\hat{\mathcal{D}}_{\mathtt{NCOMP}}$
\EndProcedure
\end{algorithmic}
\end{algorithm}

\noindent 
\begin{theorem} \label{ncomp_ach}
\textit{In Bernoulli group testing with $\alpha = \frac{\log 2}{1-q}$, under dilution noise with parameter $q \in (0,1)$, for $n$ items and $d = \Theta(n^{\theta})$ defectives, for a fixed $\theta \in (0,1)$, there holds that}
\begin{align*}
     \text{N}_{\mathtt{NCOMP}}  & \leq \:\: \frac{C \cdot d \log n \cdot (1+o(1))}{(1-q)}
\end{align*}
\textit{for a numerical constant $C$, independent of $d$ and $n$.}
\end{theorem}

\noindent
\textit{Proof.} Let us define $\text{P}^{(\text{e})}_-$ and $\text{P}^{(\text{e})}_+$ to be the probability that there exists at least one defective item which is labelled as non-defective (false negative) and at least one non-defective item which is labelled as defective (false positive), respectively. Let us thus derive conditions on the number  tests which are sufficient for $\text{P}^{(\text{e})}_-$ and $\text{P}^{(\text{e})}_+$ to vanish in the limit as $n \rightarrow \infty$.

\noindent
\textbf{False negatives:} Let $i \in \mathcal{D}$ be fixed and $\mathtt{NCOMP}(i)$ denote the classification of $i$ according to $\mathtt{NCOMP}$, i.e. positive or negative. Via a union bound over the defectives, there holds that
\begin{align}  
\text{P}^{(\text{e})}_- & \leq d \sum_{t=0}^{\text{N}}\mathbb{P}[\mathcal{G}_i = t]  \mathbb{P}[\mathtt{NCOMP}(i) = 0 \: |\: \mathcal{G}_i = t]  \nonumber\\
& = d  \sum_{t=0}^{\text{N}} \binom{\text{N}}{t}\left(\frac{\alpha}{d}\right)^t\left(1-\frac{\alpha}{d}\right)^{\text{N}-t}\mathbb{P}\left( \mathcal{P}_i^+  < t(1-q(1+\Delta))\right). \label{ncomp_1}
\end{align}

Let $\eta$ denote the probability that in a given test including item i, all defectives initially included are diluted (including item i), implying a negative test outcome. Then, the inner probability above corresponds to the probability of this event occurring in at least $q(1+\Delta)$ times the number of tests where $i$ is included
\begin{align}
\eta & = q\left (1-\frac{\alpha}{d} (1-q)\right)^{d-1} = qe^{-\alpha(1-q)} (1+o(1)). \nonumber
\end{align}
We thus bound the inner probability in  (\ref{ncomp_1}) as follows
\begin{align}
\mathbb{P}(\mathcal{P}_i^+ & < t(1-q\cdot(1+\Delta)) \nonumber \\
& =  \sum_{r = t-t(1-q(1+\Delta))}^t \binom{t}{r}\eta^r \left(1-\eta\right)^{t-r} \nonumber\\[5pt]
& = \mathbb{P}\left[\text{Bin}(t,\eta) \geq t-t(1-q(1+\Delta))\right] \nonumber \\[5pt]
& = \mathbb{P}\left[\text{Bin}(t,\eta) \geq t\eta  \left ( 1+\frac{q(1+\Delta)-\eta}{\eta}\right)\right] \nonumber \\[5pt]
& \leq \exp\left(-2t  (q(1+\Delta)-\eta)^2\right) \label{ncomp_4}
\end{align}
where (\ref{ncomp_4}) follows by applying the upper tail concentration inequality in Section \ref{appendix_chernoff} on the above $\text{Bin}(t,\eta)$ random variable, provided that $q(1+\Delta) - \eta > 0$. Inserting into (\ref{ncomp_1}) 
\begin{align}
    \text{P}^{(\text{e})}_- & \leq d \sum_{t=0}^{\text{N}} \binom{\text{N}}{t}  \left( \frac{\alpha}{d}e^{-2(q(1+\Delta)-\eta)^2}\right)^t  \left(1-\frac{\alpha}{d}\right)^{\text{N}-t} \nonumber \\
    & = d \left(1-\frac{\alpha}{d}\left(1-e^{-2(q(1+\Delta)-\eta)^2}\right)\right)^{\text{N}} \label{ncomp_6}\\[5pt]
    & \leq d  \exp\left(-\frac{\alpha}{d}\left(1-e^{-2(q(1+\Delta)-\eta)^2}\right) \text{N}\right) \label{ncomp_7}\\[5pt]
    & \leq d \exp\left(-\frac{\alpha}{d} (1-e^{-2}) (q(1+\Delta)-\eta)^2  \text{N}\right) \label{ncomp_8}
\end{align}
where (\ref{ncomp_6}) follows from the binomial theorem, (\ref{ncomp_7}) from truncation of the Taylor series of the exponential function, (\ref{ncomp_8}) by a suitable bound on $x \mapsto 1-e^{-2x}$, which holds provided that $q(1+\Delta) - \eta \in [-1,1]$. 
By combining the conditions imposed on $q(1+\Delta)$ for the concentration inequality in (\ref{ncomp_4}) to hold together with the condition required in (\ref{ncomp_8}), one obtains
\begin{align}
         \text{P}^{(\text{e})}_- &\leq  \exp\left(\log d-\frac{\alpha}{d} \left(1-e^{-2}\right)\cdot \left(q(1+\Delta)-\eta\right)^2  \text{N}\right)  \label{ncomp_9}
\end{align}
for $\eta  < q(1+\Delta) \leq \eta +1$ and $\eta = qe^{-\alpha(1-q)}\cdot(1+o(1))$. For $d = \Theta(n^{\theta})$, in order for $\text{P}^{(\text{e})}_- \rightarrow 0$, as $n \rightarrow \infty$ it suffices that 
\begin{align}
   \log d-\frac{\alpha}{d} \left(1-e^{-2}\right)\left(q(1+\Delta)-\eta\right)^2  \text{N} \leq -\gamma(n)\label{gamma_slow}
\end{align}
where $\gamma(n)$ denotes an arbitrarily slowly growing positive function of n. A sufficient condition for the probability that $\mathtt{NCOMP}$ outputs false negatives to be vanishing as $n \rightarrow \infty$ is obtained by
\begin{align}
    \text{N} \geq  \:\: \frac{d \log d \cdot (1+o(1))}{\alpha q^2  (1-e^{-2})\cdot((1+\Delta)-e^{-\alpha(1-q)})^2}\label{T_negatives}.
\end{align} 

\noindent
\textbf{False positives:} Let us fix $j \notin \mathcal{D}$. Via a union bound over the non-defective items, there holds that
\begin{align*}
\text{P}^{(\text{e})}_+ & \leq (n-d)  \sum_{t=0}^{\text{N}}\mathbb{P}[\mathcal{G}_j=t]  \mathbb{P}[\mathtt{NCOMP}(j) = 1 \: |\: \mathcal{G}_j = t]\\
\end{align*}
\begin{align}
& = (n-d)  \sum_{t=0}^{\text{N}} \binom{\text{N}}{t} \left(\frac{\alpha}{d}\right)^t   \left(1-\frac{\alpha}{d}\right)^{\text{N}-t} \label{ncomp_10} \\
& \:\:\:\:\:\:\:\:\:\:\:\:\:\:\:\:\:\:\:\:\:\:\:\:\cdot\mathbb{P}( \mathcal{P}_j^+  \geq t(1-q\cdot(1+\Delta)). \nonumber
\end{align}
\noindent
Let $\psi$ denote the probability of the event that in a given test where item $j$ is included, at least one other defective item is also included and is not diluted, hence hiding $j$ and generating a positive test result. The inner probability in (\ref{ncomp_10}) corresponds to the probability of this event occurring in at least $(1-q(1+\Delta))$ times the  number of tests where item $j$ is included. This follows from the fact that each test is independent from the others in terms of item composition and dilution of its 1-entries. We now have the following
\begin{align}
\psi & = 1-\left (1-\frac{\alpha}{d}(1-q)\right)^d \nonumber = (1-e^{-\alpha(1-q)})\cdot  (1+o(1)). \nonumber
 \end{align}
We thus bound the inner probability in (\ref{ncomp_10}) similarly
 \begin{align}
    & \mathbb{P}\left(\mathcal{P}_j^+ \geq t(1-q(1+\Delta))\right) \nonumber \\[5pt]
    & = \sum_{r = t(1-q(1+\Delta))}^t \binom{t}{z} \psi^r  (1-\psi)^{t-r} \nonumber \\[5pt]
    & = \mathbb{P}\left[\text{Bin}(t,\psi) \geq t  (1-q(1+\Delta))\right] \nonumber \\[5pt] 
    & = \mathbb{P}\left[ \text{Bin}(t,\psi)  \geq t \psi \left( 1+\frac{1-q(1+\Delta)-\psi}{\psi} \right)\right] \nonumber \\[5pt]
    & \leq \exp\left(-2t   (1-q(1+\Delta)-\psi)^2\right) \label{ncomp_13}
\end{align}where (\ref{ncomp_13}) follows by applying the lower tail concentration inequality as in Section \ref{appendix_chernoff} on the $\text{Bin}(t,\psi)$ random variable above, provided that $1-q (1+\Delta)-\psi > 0$. Let us proceed, as before, by inserting the bound obtained in (\ref{ncomp_13}) into (\ref{ncomp_10}).
\begin{align}
    \text{P}^{(\text{e})}_+ & \leq (n-d)  \sum_{t = 0}^\text{N} \binom{\text{N}}{t} \left(\frac{\alpha}{d}e^{-2(1-q(1+\Delta  ) -
    \psi)^2}\right)^t\left (1-\frac{\alpha}{d}\right)^{\text{N}-t} \nonumber\\[5pt]
    & = (n-d)\left(1-\frac{\alpha}{d} \left(1-e^{-2(\psi-1+q(1+\Delta))^2}\right)\right)^{\text{N}} \label{ncomp_15}\\[5pt]
    & \leq (n-d)  \exp\left(-\frac{\alpha}{d} \left(1-e^{-2(\psi-1+q(1+\Delta))^2}\right)  \text{N}\right) \label{ncomp_16}\\[5pt]
    & \leq (n-d)  \exp\left(-\frac{\alpha}{d}  \left(1-e^{-2}\right) (\psi-1+q(1+\Delta))^2   \text{N}\right)\label{ncomp_17}
\end{align}
As before, (\ref{ncomp_15}) follows from the binomial theorem, (\ref{ncomp_16}) from truncating the Taylor series of the exponential function, (\ref{ncomp_17}) via a suitable bound on $x \mapsto 1-e^{-2x}$, which holds provided that $\psi-1+q(1+\Delta) \in [-1,1]$. By combining the conditions on $q(1+\Delta)$ for (\ref{ncomp_13}) and (\ref{ncomp_17}) to hold, one obtains  
\begin{align}
 \text{P}^{(\text{e})}_+ & \leq \exp\biggl(\log(n-d)-\frac{\alpha}{d} \left(1-e^{-2}\right) \: \cdot \label{ncomp_18} \\
& \hspace{6em}\left(\psi-1+q(1+\Delta\right))^2  \text{N} \biggr)\nonumber
\end{align}
provided that $-\psi \leq q(1+\Delta) < 1-\psi$, for $\psi = (1-e^{-\alpha(1-q)})\cdot(1+o(1))$. For $d = \Theta(n^{\theta})$, in order for for $\text{P}^{(\text{e})}_+ \rightarrow 0$, as $n \rightarrow \infty$ it suffices that
\begin{align}
   \log(n-d)-\frac{\alpha}{d}  (1-e^{-2}) (\psi-1+q(1+\Delta))^2  \text{N} \leq -\gamma(n) \nonumber
\end{align}
where $\gamma(n)$ is chosen as in (\ref{gamma_slow}). Thus, a sufficient condition for the probability that $\mathtt{NCOMP}$ outputs false positives to vanish as $n \rightarrow \infty$ is
\begin{align}
    \text{N} & \geq   \frac{d \log (n-d) \cdot (1+o(1))} {\alpha q^2(1-e^{-2})\cdot ((1+\Delta)-\frac{1}{q}e^{-\alpha(1-q)})^2}. \label{T_positives}
\end{align}
By combining the results in (\ref{T_negatives}) and (\ref{T_positives}) with the conditions on $(1+\Delta)$, as specified in (\ref{ncomp_9}) and (\ref{ncomp_18}), the following is a sufficient condition for $\text{P}^{(\text{e})}_{\mathtt{NCOMP}} \rightarrow 0$ as $n \rightarrow \infty$
\begin{align*}
 \text{N} & \geq \max \biggl[\frac{d \log d\cdot(1+o(1))}
        {\alpha q^2  (1-e^{-2})
        \cdot ((1+\Delta)-e^{-\alpha(1-q)})^2},\\
    & \hspace{3em}\frac{d \log (n-d)  \cdot (1+o(1)) }
        {\alpha q^2(1-e^{-2})\cdot ((1+\Delta)-\frac{1}{q}e^{-\alpha(1-q)})^2}\biggr]
\end{align*}
\noindent
provided that $e^{-\alpha(1-q)} < (1+\Delta) < \frac{1}{q}e^{-\alpha(1-q)}$.  Let us note that in the above achievable bound, the terms in $\max[\cdot]$ are inverses of two parabolas viewed as functions of $1+\Delta$, with asymptotes at $e^{-\alpha(1-q)}$ and $\frac{1}{q}e^{-\alpha(1-q)}$, respectively. The two asymptotes coincide with the boundaries of the feasible region for $1+\Delta$, as above. It thus follows that the value of $1+\Delta$ which minimises the expression obtained for the achievability bound is at the intersection of these inverse parabolas, denoted by $1+\tilde{\Delta}$, for which an explicit expression is provided below.
\begin{align*}
 & 1+\tilde{\Delta} = \\
  & \frac{e^{-\alpha\left(1-q\right)}}{q}  \left(\frac{\sqrt{\log\left(d\right) \log\left(n-d\right)}  \left(1-q\right)+\log\left(\frac{\left(n-d\right)^{q}}{d}\right)}{\log\left(\frac{n-d}{d}\right)}\right)
\end{align*}
At $1+\tilde{\Delta}$, the two expressions in $\max[\cdot]$ in the achievability bound above match. Thus, a sufficient number of tests above which $\text{P}_{\mathtt{NCOMP}}^{(e)}$ vanishes in dilution noise (or equivalently an upper bound on $\text{N}_{\mathtt{NCOMP}}$) may be written as follows 
\begin{equation*}
     \text{N} \geq  \:\: \frac{d \log n \cdot (1+o(1)) }{\alpha q^2 (1-e^{-2})\cdot ((1+\tilde{\Delta})-e^{-\alpha(1-q)})^2} \geq  \:\: \text{N}_{\mathtt{NCOMP}}.
\end{equation*}Additionally, one may show that
\begin{align*}
    1+\tilde{\Delta} = \frac{e^{-\alpha(1-q)}}{q}\frac{\sqrt{\theta}(1-q)+q-\theta}{1-\theta}\cdot (1+o(1)).
\end{align*}
For $\alpha = \frac{\log 2}{1-q}$, the above bound on $\text{N}_{\mathtt{NCOMP}}$ reduces to the required bound in \textit{Theorem \ref{ncomp_ach}}. As we will show in the next section, this choice for $\alpha$ turns out to be optimal in terms of matching achievability and converse bounds in the limit as $n \rightarrow\infty, q\rightarrow 1$. This completes the proof. \qedsymbol

\vspace{1em}
Based on the arguments above, we now provide the corresponding achievable rate for $\mathtt{NCOMP}$. Using the asymptotic expression for rates as defined in Section \ref{setup} and based on \textit{Theorem \ref{ncomp_ach}}, an achievable rate for $\mathtt{NCOMP}$ is provided by
\begin{align*}
    \text{R}_{\mathtt{NCOMP}}(q) \sim \frac{4(1-\theta)\cdot (1-q)}{(1-e^{-2})\log 2} \in \Omega((1-\theta)\cdot(1-q)).
\end{align*}

\section{Converse Bounds }
We now provide a converse bound valid for any choice of decoder. We denote the $\text{Poisson}(\alpha)$ distribution as $\text{P}(\alpha)$. 

\begin{theorem} \label{strong_conv}
\textit{In Bernoulli group testing with $\alpha = \frac{\log 2}{1-q}$, under dilution noise with parameter $q \in (0,1)$, for $n$ items and $d = \Theta(n^{\theta})$ defectives, for a fixed $\theta \in (0,1)$ and for any decoder $\mathtt{A}$, there holds that }
    \begin{align*}
        \text{N}_{\mathtt{A}} \geq \frac{\log_2\binom{n}{d}\cdot (1+o(1))}{1 - \mathbb{E}_{Z \sim \text{P}(\alpha)}\left[H_b\left(q^Z\right)\right]}
    \end{align*}
\textit{where $\text{H}_b(\rho) = \rho\log_2\left(\frac{1}{\rho}\right)+(1-\rho)\cdot \log_2\left(\frac{1}{1-\rho}\right)$ is the binary entropy function.}
\end{theorem}

\noindent
\textit{Proof.} Following expression (4.30) in \cite{aldridge2019group}, we deduce from Fano's inequality that for any decoder $\mathtt{A}$ achieving vanishing error probability, it is necessary that 
\begin{align} \label{fano}
    \text{N}_{\mathtt{A}} \geq \frac{\log_2{n\choose{d}}\cdot (1+o(1)) }{\mathcal{I}(\text{M}_{i,\cdot}; \mathbf{y}_i)} 
\end{align}
where $\text{M}_{i,\cdot}$ denotes the $i$-th row of the i.i.d. Bernoulli test matrix (a test vector whose entries have probability $\frac{\alpha}{d}$ of being equal to 1), whilst $\mathbf{y}_{i}$ is the Boolean valued outcome of the corresponding test. We rewrite the mutual information as $\mathcal{I}\left(\text{M}_{i,\cdot};\mathbf{y}_{i}\right) = \text{H}\left(\mathbf{y}_{i}\right) - \text{H}\left(\mathbf{y}_{i} \:|\: \text{M}_{i,\cdot}\right)$ and, letting $n_{\mathtt{def}}$ denote the number of defectives in a test, for $\alpha\in o(n)$ considered fixed, expand terms separately
\begin{align*}
    \text{H}\left(\mathbf{y}_{i}\right) & = \text{H}_b\left( \sum_{j = 0}^{d} \binom{d}{j}\left(\frac{\alpha}{d}q\right)^j \left(1-\frac{\alpha}{d}\right)^{d-j} \right) \\
    &  \xrightarrow{n\rightarrow \infty} \text{H}_b\left(e^{-\alpha(1-q)}\right) \\
    \text{H}\left(\mathbf{y}_{i} \:|\: \text{M}_{i,\cdot}\right) &= \sum_{j = 0}^d \mathbb{P}\left( n_{\mathtt{def}} = j \right)\cdot \text{H}\left( \mathbf{y}_{i}\:|\: n_{\mathtt{def}} = j\right) \\[5pt]
    & = \sum_{j = 0}^d \binom{d}{j}  \left(\frac{\alpha}{d}\right)^j   \left(1-\frac{\alpha}{d}\right)^{d-j}   \text{H}_b\left(q^j\right) \\
    &\xrightarrow{n\rightarrow \infty} \sum_{j = 0}^{\infty} \text{H}_b\left(q^j\right)\cdot  \frac{\alpha^j e^{-\alpha}}{j!} = \mathbb{E}_{\text{Z}\sim \text{P}(\alpha)}\left[\text{H}_b\left(q^{\text{Z}}\right)\right].
\end{align*}
Where in the second set of equations, passing the limit as $n \rightarrow \infty$ is justified due to Portmanteau's theorem (Lemma 18.9 in \cite{asymptoticstatistics}), the Binomial distribution converging weakly to the Poisson distribution, and $\text{H}_b(\cdot)$ being continuous and bounded. For $\alpha = \frac{\log{2}}{1-q}$ in the above (which as noted previously is optimal for high noise regimes in terms of matching converse and achievability bounds) and substituting them into the mutual information, (\ref{fano}) yields the required result. 
\qedsymbol

\vspace{1em}
The associated converse rate is obtained as outlined in Section \ref{setup} by $\text{R}_{\mathtt{IT}}(q) \sim 1 - \mathbb{E}_{\text{Z} \sim \text{P}(\alpha)}\left[\text{H}_b\left(q^{\text{Z}}\right)\right]$. We now claim $\lim_{q \rightarrow 1} \text{R}_{\mathtt{IT}}(q) \in \mathcal{O}(1-q)$. \\ 
Begin by noting that $\text{R}_{\mathtt{IT}}(q) \geq 0$, $\text{H}_b(x) \geq 1-4(x-\frac{1}{2})^2$, and $0 \leq \text{H}_b(x) \leq 1$, for $x \in [0,1]$. We may then provide the following upper bound on $\text{R}_{\mathtt{IT}}$
\begin{align}\label{aim_r_it}
 \text{R}_{\mathtt{IT}}(q) & \leq 4 \cdot \mathbb{E}_{\text{Z}\sim \text{P}(k\log2)}\left [ \min \left \{ \left( q^Z -\frac{1}{2}\right )^2, \frac{1}{4}\right \} \right ].
\end{align}
For ease of exposition, let $q = 1-\frac{1}{k}$, where k scales to infinity as $q \rightarrow 1$, and write $Z = k \log 2 + \epsilon$. We will show that $\text{R}_{\mathtt{IT}} \lesssim \frac{1}{k}$, where the notation $\lesssim$ is used to indicate the omission of numerical constants from the numerator on the right side. \\
Using the fact that $-\frac{1}{k-1} \leq \log\left(1-\frac{1}{k}\right) \leq -\frac{1}{k}$ for positive $k$ and taking logarithms, we obtain
\begin{align*}
    \left|\left(1-\frac{1}{k}\right)^{k\log 2 + \epsilon}-\frac{1}{2}\right| & \leq \left| \frac{1}{2}- \exp\left(-\log 2-\frac{\epsilon+\log 2}{k-1}\right)\right| \\
    & \leq \frac{\left|\epsilon\right|+\log 2}{k-1}
\end{align*}
for $\left|\epsilon\right| \leq k-2$, where $\left| \frac{1}{2}- \exp\left(-\log 2-\delta\right)\right| \leq \left | \delta \right|$ for any $\left|\delta\right| \leq 1$. 
For $f(\epsilon) := \left(1-\frac{1}{k}\right)^{k\log 2 + \epsilon}-\frac{1}{2}$, we rewrite (\ref{aim_r_it}) as 
\begin{align*}
    \text{R}_{\mathtt{IT}}(q) & \leq \mathbb{E}_{\epsilon}\left[\min \left \{ 4f(\epsilon)^2,1  \right\}\right]
\end{align*}
where for $\left|\epsilon\right| \leq k-2$, it holds that $\left|f(\epsilon) \right| \leq \frac{\left| \epsilon\right| + \log 2}{k-1}$. \\ 
For $C$ constant, we bound the above expectation as follows
\begin{align}
    \text{R}_{\mathtt{IT}}(q) & \leq \sum_{n = 0}^{\infty} \mathbb{E}_{\epsilon} \left[\min \left\{ 4f(\epsilon)^2,1  \right\} | \:\: n \leq \frac{\left|\epsilon\right|}{C\sqrt{k}} \leq n+1 \right] \nonumber\\ 
    & \hspace{7em}\cdot \mathbb{P}\left(n \leq \frac{\left|\epsilon\right|}{C\sqrt{k}} \leq n+1\right) \\
     & \leq \sum_{n = 0}^{\floor{k^{\frac{1}{4}}}} \: \left(\underset{\frac{\left|\epsilon\right|}{C\sqrt{k}} \leq n+1}{\max} 4f(\epsilon)^2\right) \mathbb{P}\left(\left| \epsilon \right| \geq nC \sqrt{k}\right) \nonumber\\
     & \hspace{7em}+ \mathbb{P}\left(\frac{\epsilon}{C\sqrt{k}} \geq \floor{k^{\frac{1}{4}}}\right)\label{last}
\end{align}
where (\ref{last}) follows by choosing $k$ large enough such that $\frac{\left|\epsilon\right|}{C\sqrt{k}} \leq \floor{k^{\frac{1}{4}}} + 1$, which guarantees that $\left| \epsilon\right| \leq k-2$. Using Poisson concentration inequalities as in \cite{poissonconc} and noting that $C \lesssim 1, n \lesssim k^{\frac{1}{4}}, 2Cn\sqrt{k} \ll 2k\log 2, 2\log 2 < 2$, one obtains
\begin{align}
\mathbb{P}\left(\left|\epsilon\right| \geq nC\sqrt{k} \right)
& \leq 2\exp\left(\frac{-Cn^2}{2}\right). \label{step1}
\end{align}
In addition, we note that for $\left| \epsilon \right| \leq k - 2$ (as in the summation term in the upper half of (\ref{last})), one may write $4f(\epsilon)^2 \leq 4\left(\frac{\left|\epsilon\right| + \log 2}{k-1}\right)^2 \label{step3}$. Combining this with (\ref{last}) and (\ref{step1}), we obtain
\begin{align}
    & \text{R}_{\mathtt{IT}}(q) \leq \nonumber\\
    & \sum_{n = 0}^{\floor{k^{\frac{1}{4}}}} \left(4\frac{(n+1)C\sqrt{k} + \log 2}{k-1}\right)^2 \cdot \exp\left(\frac{-Cn^2}{2}\right) + o\left(\frac{1}{k}\right) \nonumber \\
    & \leq \frac{5}{k}\sum_{n = 0}^{\infty}(n+1)^2\exp\left(-\frac{n^2}{2}\right) \nonumber \lesssim \frac{1}{k} \nonumber
\end{align}
which yields the required result.

\section{Discussion}
In this work we have analyzed the Bernoulli group testing problem under dilution noise and showed that the optimal choice of average test sparsity depends on the level of dilution noise, producing tighter matching achievability and converse bounds up to order for high noise levels. We highlight that our achievability result from \textit{Theorem 1} scales as $\mathcal{O}(\frac{d \log n}{1-q})$ as opposed to the $\mathcal{O}(\frac{d \log{n}}{(1 - q)^2})$ scaling in \cite{atia2012boolean} with fixed $\alpha$, matching our converse result scaling at $\Omega(\frac{d \log n}{1-q})$ as $q \rightarrow 1$. 
Several interesting phenomena emerge when considering dilution of sampled items from test pools, and in this work we demonstrate that a noise-level-dependent test sparsity parameter has the qualitative effect of offsetting the dilution noise distribution over the test vector. This in turn produces tight achievability and converse bounds for high noise regimes, a feature not observed in other well-studied noise models such as the Z-channel. 
Moreover, we formulate a question regarding the full information-theoretic consequences of dilution noise by considering the case in which a test is repeated multiple times. Whilst in the noiseless case one retrieves information solely on the presence of a defective in the given pool of items, in the dilution model the distribution of the test results carries additional information on the number of defectives. This indicates that the dilution channel may provide more information about the defective set than the noiseless one in the above instance. Mathematically certifying that dilution noise is disadvantageous in information-theoretic terms within Bernoulli group testing is left as an open question,  together with further investigations of the broader role that dilution noise plays in sparse inference.

\section{Appendix}\label{appendix_chernoff}
The following tail bounds for binomial random variables are used throughout the paper. For $X \sim \text{Bin}(K,p)$, $\mu := \mathbb{E}[X]$ and $\delta > 0$, there holds that 
\begin{align}
    \mathbb{P}[X \geq (1+\delta)\mu] & \leq \exp\left(-\frac{2 \delta^2 \mu^2}{K}\right)  \\
   \mathbb{P}[X \leq (1-\delta)\mu] &\leq \exp\left(-\frac{\delta^2 \mu^2}{K}\right) \label{weak_conc}
\end{align}

\section{Acknowledgments}
The authors would like to thank George Atia, Oliver Gebhard, Venkatesh Saligrama, Jonathan Scarlett, and Pedro Abdalla Teixeira for helpful discussions on the group testing problem.

\bibliography{bibliography.bib}
\bibliographystyle{abbrv}

\end{document}